\documentclass[structabstract]{aa}  
\usepackage{amstext,amsmath}
\usepackage{graphicx}
\usepackage{textcomp}
\usepackage{txfonts}
\usepackage{wasysym}
\usepackage{time}
\usepackage{subfig}
\usepackage{color}
\usepackage{url}
\usepackage{natbib}
\bibpunct{(}{)}{;}{a}{}{,}

\usepackage{multirow} 
\usepackage{array} 
\usepackage{booktabs}
\usepackage{balance}

\begin{document}
\title{Sources of straylight in the post-focus imaging instrumentation of the
  Swedish 1-m Solar Telescope}

\author{M.G.\ L{\"o}fdahl\inst{1,2} \and G.B. Scharmer\inst{1,2}}
\institute{Institute for Solar Physics, Royal Swedish Academy of
  Sciences, Albanova University Center, 106\,91 Stockholm, Sweden \and
  Stockholm Observatory, Dept. of Astronomy, Stockholm University,
  Albanova University Center, 106\,91 Stockholm, Sweden}
\authorrunning{L{\"o}fdahl \& Scharmer}

 \date{Received August 30, 2011; accepted September 30, 2011}
\frenchspacing

\abstract
{Recently measured straylight point spread functions (PSFs) in
  Hinode/SOT make granulation contrast in observed data and synthetic
  magnetohydrodynamic (MHD) data consistent. Data from earthbound
  telescopes also need accurate correction for straylight and fixed
  optical aberrations.}
{We aim to develop a method for measuring straylight in the post-focus
  imaging optics of the Swedish 1-m Solar Telescope (SST).}
{We removed any influence from atmospheric turbulence and scattering
  by using an artificial target. We measured integrated straylight
  from three different sources in the same data: ghost images caused
  by reflections in the near-detector optics, PSFs corresponding to
  wavefront aberrations in the optics by using phase diversity, and
  extended scattering PSF wings of unknown origin by fitting to a
  number of different kernels. We performed the analysis separately in
  the red beam and the blue beam.}
{Wavefront aberrations, which possibly originate in the bimorph mirror
  of the adaptive optics, are responsible for a wavelength-dependent
  straylight of 20--30\% of the intensity in the form of PSFs with
  90\% of the energy contained within a radius of 0\farcs6. There are
  ghost images that contribute at the most a few percent of
  straylight. The fraction of other sources of scattered light from
  the post-focus instrumentation of the SST is only $\sim$$10^{-3}$ of
  the recorded intensity. This contribution has wide wings with a FWHM
  $\sim$16\arcsec{} in the blue and $\sim$34\arcsec{} in the red.}
{The present method seems to work well for separately estimating
  wavefront aberrations and the scattering kernel shape and fraction.
  Ghost images can be expected to remain at the same level for solar
  observations. The high-order wavefront aberrations possibly caused
  by the AO bimorph mirror dominate the measured straylight but are
  likely to change when imaging the Sun. We can therefore make no
  firm statements about the origin of straylight in SST data, but
  strongly suspect wavefront aberrations to be the dominant source.}

\maketitle

\section{Introduction}
\label{sec:introduction}

For many years, there has been a discrepancy between the contrast in
observed solar images and the corresponding images produced by
magnetohydrodynamic (MHD) simulations. It was not clear whether
compensation for straylight in the observations or missing physics in
the simulations were at fault. This has now been resolved by
measurements of the point spread function (PSF) in the Solar Optical
Telescope (SOT) on the Hinode spacecraft \citep{2008a&a...487..399w}
and comparison with artificial data \citep{2009a&a...503..225w}.
\citet[see their introduction for a full account of the
problem]{scharmer10high-order} recently initiated a project to measure
the straylight sources of the Swedish 1-meter Solar Telescope
\citep[SST; ][]{scharmer03new} to also correct SST images for the
missing contrast.

\citet{scharmer10high-order} found that a significant part of the
contrast reduction can be explained by high-order modes in the pupil
wavefront phase that are not corrected by the Adaptive Optics
\citep[AO; ][]{scharmer03adaptive} or image restorations with
multi-frame blind deconvolution \citep[MFBD; ][]{lofdahl02multi-frame}
techniques because of the finite number of modes used in those
techniques. The authors also implemented a method for correcting the
contrast by means of the known statistics of atmospheric turbulence
and simultaneous measurements of Fried's parameter $r_0$ from a
wide-field wavefront sensor. However, the contrast correction they
found is not sufficient to reach the contrasts expected from MHD
simulations and Hinode observations.

In this paper we continue the search for the missing contrast by
examining the post-focus optics of the SST. By inserting targets in
the Schupmann focus, which is located close to the exit of the vacuum
system, we removed all effects upstream of that point (atmosphere and
telescope) and considered only error sources on the optical table. We
estimated the wavefront PSF by using phase diversity \citep[PD;
][]{gonsalves79wavefront,paxman92joint,lofdahl94wavefront} and fitted
the remaining straylight to a scattering PSF. In particular, we were
interested in the integrated fraction of scattered light in the image
data and the width of the scattering kernel.

\section{Imaging model}

\begin{table*}[!t]
  \centering
  \caption{Scattering kernels}
  \label{tab:kernels}
  \begin{tabular}{llll}
    \hline
    \hline\noalign{\smallskip}
    Kernel name & Definition  & FWHM \\
    \hline\noalign{\smallskip}
    Gauss & 
    $ K_\text{G}(r;\sigma)  \propto \exp(-r^2/(2\sigma^2))$ & 
    $W_\text{G} = \sigma\cdot 2 \sqrt{2 \ln 2}$ \\\noalign{\smallskip}
    Lorentz & 
    $K_\text{L}(r;\gamma)  \propto (1+r^2/\gamma^2 )^{-1}$ &
    $W_\text{L} = \gamma\cdot 2$ \\\noalign{\smallskip}
    Moffat & 
    $K_\text{M}(r;\alpha,\beta) \propto (1+r^2/\alpha^2)^{-\beta}$ & 
    $W_\text{M} = \alpha\cdot 2\sqrt{2^{1/\beta}-1}$\\\noalign{\smallskip}
    Voigt & 
    $K_\text{V}(r;\sigma,\gamma)\propto K_\text{G}(r;\sigma) * K_\text{L}(r; \gamma)$ &
    $W_\mathrm{V}\approx 0.5346 W_\mathrm{L}+\sqrt{0.2166 W_\mathrm{L}^2+W_\mathrm{G}^2}$\\ 
    \hline
  \end{tabular}
  \tablefoot{All kernels are functions of the radial coordinate
    $r=(x^2+y^2)^{1/2}$. They are normalized in the Fourier domain by 
    division with the value in the origin. $K_\text{M}$ is from
    \citet{1969A&A.....3..455M}. The approximate expression for
    $W_\mathrm{V}$ is given by \citet{olivero77empirical} and is
    claimed to be accurate to within a few \textperthousand.}
\end{table*}

We model the image formation as
\begin{equation}
  g = f * s_\phi * s_K + d + n
  \label{eq:1}
\end{equation}
where $g$ is a flat-fielded and dark-corrected data frame, $f$ is the
object, $s_{\phi}$ is the point-spread function (PSF) of wavefront
errors $\phi$, $s_K$ is a scattering PSF with extended wings,
$d$ represents a residual dark level that was not subtracted properly
in flat-fielding, and $n$ is additive Gaussian white noise. These
quantities are all functions of the spatial coordinates $(x,y)$,
suppressed when possible for compact notation. The symbol $*$ denotes
convolution.

The scattering PSF, $s_K$, is modeled as
\begin{equation}
  s_K = c \delta +(1-c)\cdot K
  \label{eq:2}
\end{equation}
where $\delta$ is the Dirac delta function, $K$ is a convolution
kernel equal to one of $K_\text{M}$, $K_\text{L}$, $K_\text{G}$, or
$K_\text{V}$ as given in Table~\ref{tab:kernels}. Because it is part
of the scattering PSF, we will refer to $K$ as a scattering kernel.
Both the $\delta$ function and the kernels are normalized in the
Fourier domain by division with the value in the origin. Owing to this
normalization, $(1-c)$ expresses the integrated fraction of this type
of straylight.

\section{Data}
\label{sec:29maydata}

The data used in this experiment were collected with the SST
\citep{scharmer03new} on 29 May 2010 between 17:10 and 18:40 UT.

\begin{figure}[!t]
  \centering
  \hfill
  \begin{minipage}[c]{0.35\linewidth}
    \includegraphics[bb=222 166 276 226,angle=-90,clip,width=\linewidth]{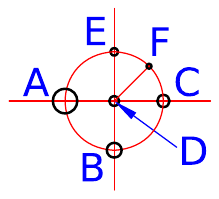}\\[2mm]
  \end{minipage}
  \quad
  \begin{minipage}[c]{0.55\linewidth}
    \vspace{-4mm}
    \begin{tabular}{lrrr}
      \hline\hline\noalign{\smallskip}
      \multirow2*{Hole} & $x$  & $y$ & \diameter\\
      &(mm)&(mm)&(\textmu{}m)\\
      \hline\noalign{\smallskip}
      \textsf{A}& $-2$ & $ 0$ & $1000$\\
      \textsf{B}& $ 0$ & $-2$ &  $250$\\
      \textsf{C}& $ 2$ & $ 0$ &  $120$\\
      \textsf{D}& $ 0$ & $ 0$ &   $60$\\
      \textsf{E}& $ 0$ & $ 2$ &   $30$\\
      \textsf{F}& $\sqrt 2$ & $\sqrt 2$ & $20$ \\
      \hline
    \end{tabular}
  \end{minipage}
  \hfill
  \caption{Straylight target drawing. Positions $(x,y)$ in mm,
    diameters (\diameter) in \textmu{}m with tolerances of
    1--2~\textmu{}m.}
  \label{fig:target-drawing}
\end{figure} 

The primary optical system of the SST is a singlet lens with a focal
length of 20.3~m at 460~nm and a 98-cm aperture. A mirror at the
primary focus reflects the light to a Schupmann corrector, which forms
an achromatic focus next to the primary focus. In this focal plane we
placed an artificial target to isolate the optical aberrations and
scattering downstream from this point. The target, manufactured by
Molenaar Optics, is a 25~\textmu{}m thick metal foil with six holes of
different sizes, see Fig.~\ref{fig:target-drawing}. The manufacturing
tolerances are very small, but close inspection of the images revealed
small irregularities at the edges of the larger holes, probably caused
by dust particles. Using a motor stage, we were also able insert a
pinhole array used mainly for alignment purposes.

\begin{figure}[!t]
  \centering
  \includegraphics[bb=138 420 412 658,width=\linewidth]{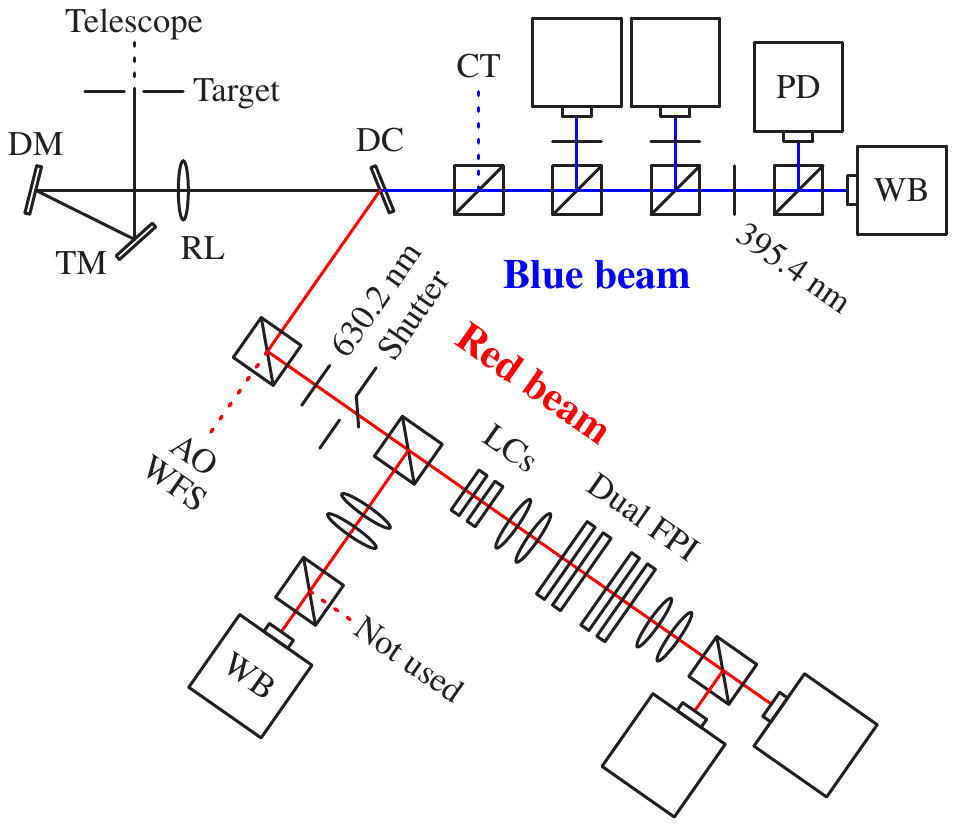}
  \caption{Setup schematics. Light from the telescope enters through
    the target from upper left. TM = tip-tilt mirror; DM = deformable
    mirror; RL = reimaging lens; DC = dichrocic beamsplitter; CT =
    correlation tracker; WFS = wavefront sensor; LCs = Liquid
    Crystals; FPI = Fabry--P\'erot interferometer. Only the cameras
    that produced images for this investigation are labeled (WB, PD).
    The figure is not to scale and the following intentional errors
    are introduced to save space: the angle to the beam reflected off
    the DC is drawn too wide and the transmitted and reflected beams
    from the first red beamsplitter are switched.}
  \label{fig:setup}
\end{figure}

The setup following the Schupmann focus is illustrated in
Fig.~\ref{fig:setup}. The beam expands (via the tip-tilt mirror) to a
pupil plane at the location of the bimorph deformable mirror (DM).
Here, the telescope pupil is re-imaged by a field lens located just in
front of the Schupmann focus. The pupil diameter is 34~mm at this
location. There is a 35~mm diameter pupil stop at the DM. This defines
the pupil for pinhole images. The reimaging lens then makes a F/46
beam parallel to the optical table.
The light is then split by the 500~nm dichroic beamsplitter into a
blue beam and a red beam. Both beams have several cameras behind
different filters. Table~\ref{tab:setup} summarizes the cameras and
setup parameters used for this experiment.

\begin{table}[!t]
  \centering
  \caption{Setup in the two beams.}
  \begin{tabular}{lll}
    \hline
    \hline\noalign{\smallskip}
    Item & Blue beam & Red beam\\
    \hline\noalign{\smallskip}
    Wavelength (nm) &  395.4 & 630.2 \\
    No. of cameras & 4 & 3\\
    No. of cameras used & 2 & 1\\
    Cameras &\tiny MegaPlus II es4020 &\tiny Sarnoff CAM1M100\\
    FOV (pixels) & 2048$\times$2048& 1024$\times$1024\\
    Image scale (arcsec/pix) & 0.034 & 0.059\\
    Exposure time (ms) & 10  & 17  \\
    \hline
  \end{tabular}
  \label{tab:setup}
\end{table}

In the blue beam one camera was nominally a ``focus'' camera and the
other a defocused ``diversity'' camera of a PD pair. These two cameras
and their beamsplitter were mounted on a common holder and could be
moved together along the optical axis. This made it possible to
generate additional focus diversities without changing the relative
diversity between the wideband (WB) and PD cameras. The holder and its
cameras are covered by a box that blocks light from directions other
than the beam.
The relative rotation of the field in the WB and PD cameras was
measured by comparing the grid patterns of the pinhole array images.
The rotational misalignment is smaller than 0\fdg1.

In the red beam we used a single camera that was also covered by a box
to block external straylight. Here, we could only add diversities by
moving the camera.

The AO was running in closed loop on the central pinhole (\textsf{D})
of the straylight target while the artificial target data were
collected, so aberrations are assumed to be stable during the data
collection.\footnote{Because the target was repeatedly removed from
  the beam (when collecting flat fields) and inserted (when collecting
  target data), the mirror was alternately heated by the sunlight and
  allowed to cool again. Because these thermal variations cause
  changes in the mirror's shape and stress, so the level to which the
  assumption is valid is difficult to predict.} The telescope pointing
was moving near disk center to average out the granulation structure
as well as possible. Many exposures were collected at every position
with all cameras, 3500 exposures in the blue and 10000 in the red, for
the target data as well as for dark frames and flat fields. The high
number of frames was necessary because we needed a high
signal-to-noise ratio (SNR) to measure the very faint straylight far
away from the pinholes. The target data were corrected for bias and
gain variations. The corrected image $g_i$ is calculated as
\begin{equation}
  g_i = (h_i-d)/(f-d),
  \label{eq:flat}
\end{equation}
where $h_i$ is the observed data frame, $d$ is the average dark frame,
and $f$ is the average flat field image. To minimize the influence
from solar features in the averaged flat-field images, the individual
frames were collected while the telescope was circled over a quiet
area near disk center and the DM produced random wavefronts.

The images collected in different focus positions were not recorded
simultaneously and we took steps to ensure that tiny image movements
did not violate the assumption that we made in PD processing, i.e.
that there is a common object in all images after summation.
In a first pass, the data frames $g_i$ were co-added in their original
form. In a second pass, each exposure was aligned to the summed WB
image at nominal focus (WB0 image) from the first pass (separately in
blue and red) with sub-pixel precision through centroiding (center of
mass) on one of the holes, and a new sum was formed. By using the same
image as an alignment target, both the individual images in each sum
and also the sums at different focus positions were well aligned.

\begin{figure}[!t]
  \centering
  \def\bluedir{/scratch/mats/2010.05.29/BLUE/straylight}
  \def\reddir{/scratch/mats/2010.05.29/RED/straylight}
  \def\tilewidth{0.49\linewidth}
  \reflectbox{\includegraphics[width=\tilewidth]{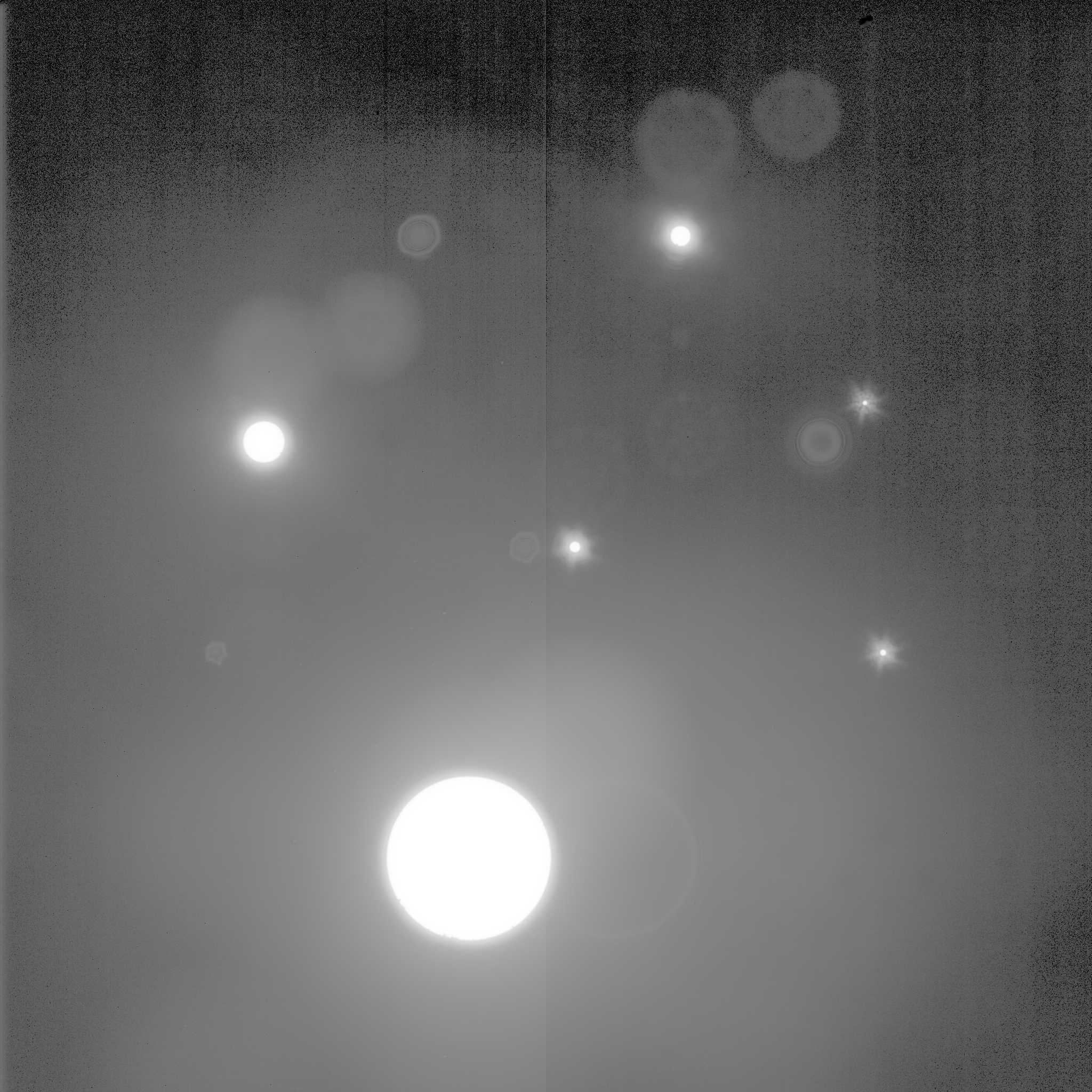}}\hfill
  \reflectbox{\includegraphics[width=\tilewidth]{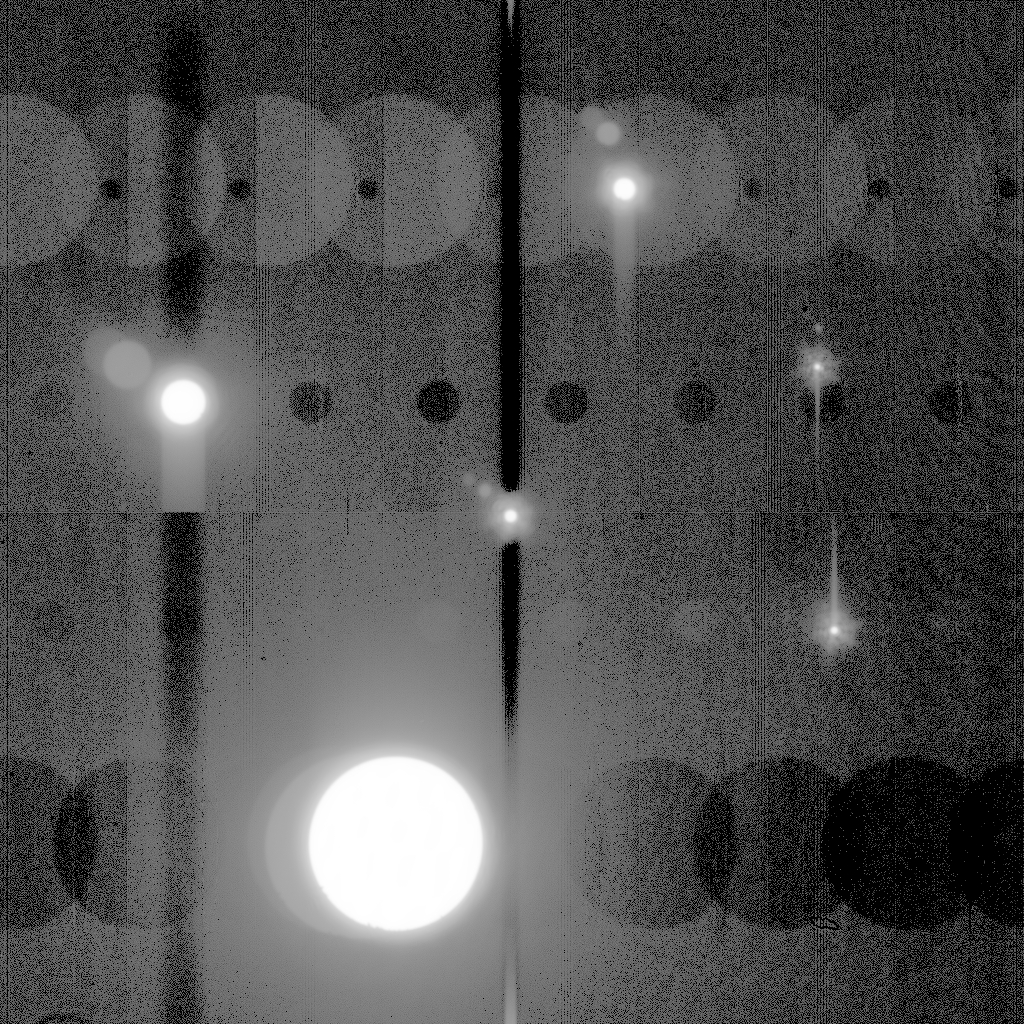}}
  \caption{Straylight target images, WB0 in both beams, displayed with
    the same log scale. \textit{Left}: blue beam; \textit{right}: red
    beam. Compare Fig.~\ref{fig:target-drawing}.}
  \label{fig:images}
\end{figure}

In the remainder of this paper we will refer to these summed images
simply as the images, corresponding to $g$ in Eq.~(\ref{eq:1}).
The WB0 images are shown in Fig.~\ref{fig:images}.

\section{Straylight measurements}
\label{sec:strayl-meas}

\subsection{Ghost images}
\label{sec:ghost-images}

Figure~\ref{fig:images} shows a variety of weak duplicates of the
pinhole images, commonly referred to as ghost images. The most
significant ghost images are shifted by only a few arcseconds and
appear to be at different focus positions. These contributions must
come from reflections in the beam splitters and other optics within a
few cm from the detector focus, and should therefore be present during
ordinary observations of the Sun as well. There are many weaker
contributions that appear to be mirror images in the blue and
repetitions in the different taps of the red CCD, probably originating
in the camera electronics.\footnote{The Sarnoff CAM1M100 CCD is
  organized in 2~by~8 taps (subfields), that are read out in
  parallel.}

The ghost images are not included in the image formation model of
Eq.~(\ref{eq:1}), but we can measure them here and then mask them in
later processing. The strongest more or less in-focus ghosts are
$\la1$\% in both the blue and red (measured on hole \textsf{A}).

\subsection{Wavefront aberrations}
\label{sec:wavefr-aberr}

\begin{figure*}[!t]
  \centering
  \def\figwidth{0.12\linewidth}
  \def\figwidth{19.2mm}
  \def\lf{0.3mm}
  \subfloat[Blue WB$-$\label{fig:blue-restored-WBm-KL}]{
    \begin{minipage}[t]{\figwidth}
      \includegraphics[width=\linewidth]{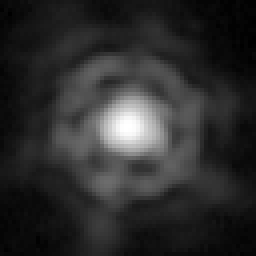}\\[\lf]
      \includegraphics[width=\linewidth]{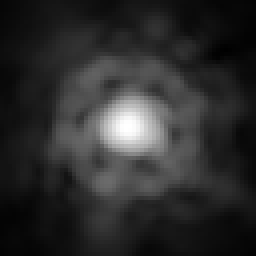}\\[\lf]
      \includegraphics[width=\linewidth]{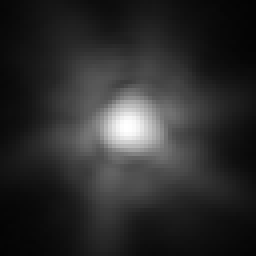}
    \end{minipage}
  }\!\!\!
  \subfloat[Blue WB$0$\label{fig:blue-restored-WB0-KL}]{
    \begin{minipage}[t]{\figwidth}
      \includegraphics[width=\linewidth]{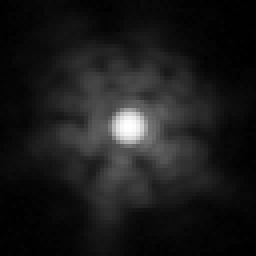}\\[\lf]
      \includegraphics[width=\linewidth]{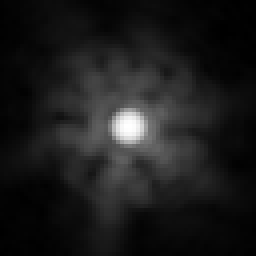}\\[\lf]
      \includegraphics[width=\linewidth]{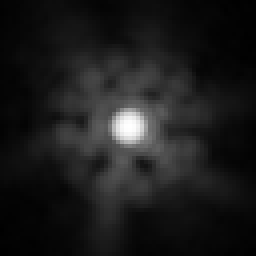}
    \end{minipage}
  }\!\!\!
  \subfloat[Blue WB$+$\label{fig:blue-restored-WBp-KL}]{
    \begin{minipage}[t]{\figwidth}
      \includegraphics[width=\linewidth]{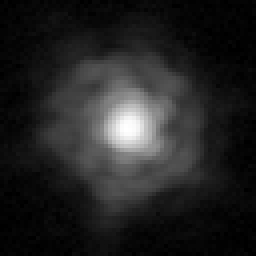}\\[\lf]
      \includegraphics[width=\linewidth]{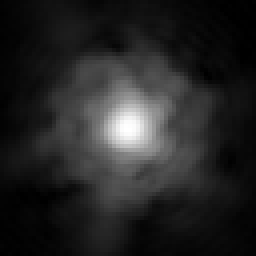}\\[\lf]
      \includegraphics[width=\linewidth]{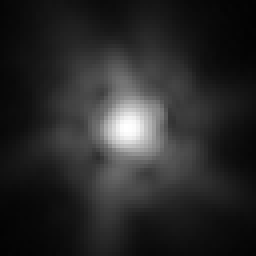}
    \end{minipage}
  }\!\!\!
  \subfloat[Blue PD$-$\label{fig:blue-restored-PDm-KL}]{
    \begin{minipage}[t]{\figwidth}
      \includegraphics[width=\linewidth]{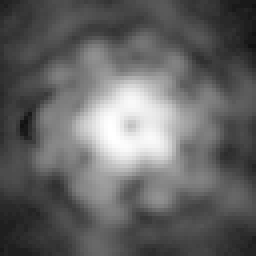}\\[\lf]
      \includegraphics[width=\linewidth]{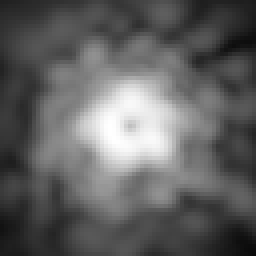}\\[\lf]
      \includegraphics[width=\linewidth]{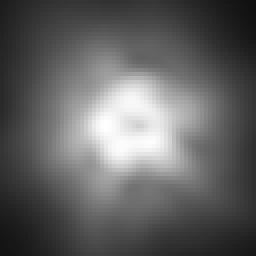}
    \end{minipage}
  }\!\!\!
  \subfloat[Blue PD$0$\label{fig:blue-restored-PD0-KL}]{
    \begin{minipage}[t]{\figwidth}
      \includegraphics[width=\linewidth]{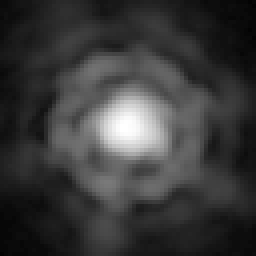}\\[\lf]
      \includegraphics[width=\linewidth]{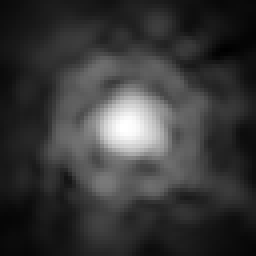}\\[\lf]
      \includegraphics[width=\linewidth]{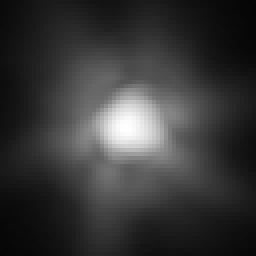}
    \end{minipage}
  }\!\!\!
  \subfloat[Blue PD$+$\label{fig:blue-restored-PDp-KL}]{
    \begin{minipage}[t]{\figwidth}
      \includegraphics[width=\linewidth]{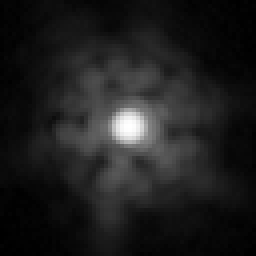}\\[\lf]
      \includegraphics[width=\linewidth]{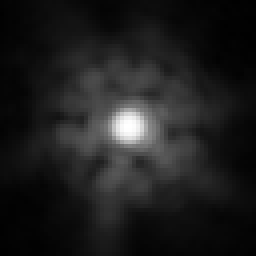}\\[\lf]
      \includegraphics[width=\linewidth]{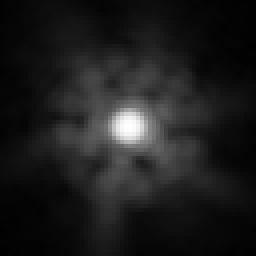}
    \end{minipage}
  }
  \subfloat[Red WB$-$\label{fig:red-restored-WBm-KL}]{
    \begin{minipage}[t]{\figwidth}
      \includegraphics[width=\linewidth]{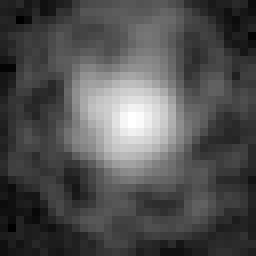}\\[\lf]
      \includegraphics[width=\linewidth]{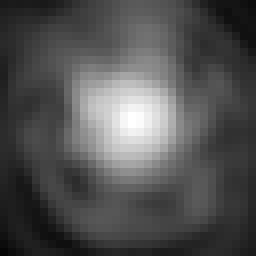}\\[\lf]
      \includegraphics[width=\linewidth]{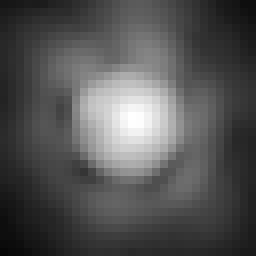}
    \end{minipage}
  }\!\!\!
  \subfloat[Red WB$0$\label{fig:red-restored-WB0-KL}]{
    \begin{minipage}[t]{\figwidth}
      \includegraphics[width=\linewidth]{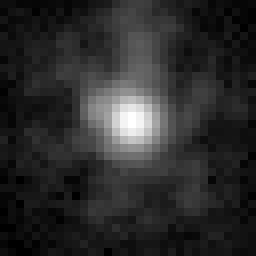}\\[\lf]
      \includegraphics[width=\linewidth]{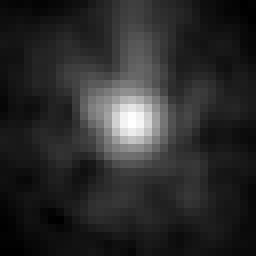}\\[\lf]
      \includegraphics[width=\linewidth]{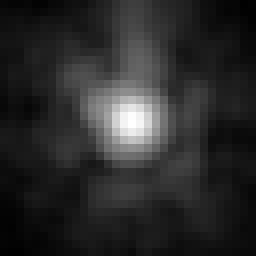}
    \end{minipage}
  }\!\!\!
  \subfloat[Red WB$+$\label{fig:red-restored-WBp-KL}]{
    \begin{minipage}[t]{\figwidth}
      \includegraphics[width=\linewidth]{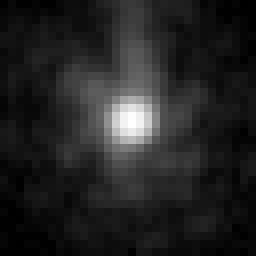}\\[\lf]
      \includegraphics[width=\linewidth]{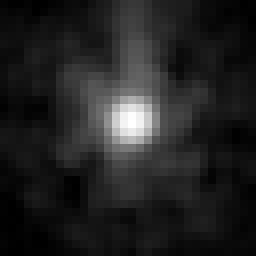}\\[\lf]
      \includegraphics[width=\linewidth]{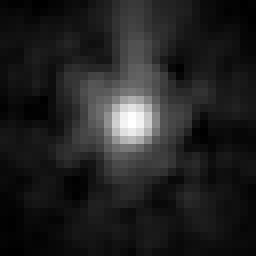}
    \end{minipage}
  }
  \caption{PD results, the central parts of target hole \textsf{F}
    images in log scale. Top row: observed image; center row:
    recreated images $M=231$; bottom row: recreated images $M=36$. The
    FOV shown is 2\farcs2$\times$2\farcs2 in the blue and
    1\farcs9$\times$1\farcs9 in the red. Beam, camera and diversity as
    noted in subcaptions a)--i).}
  \label{fig:restored}
\end{figure*}

We estimated the wavefront error PSF, $s_\phi$, through PD with the
MOMFBD C++ code of \citet{noort05solar}. We processed a subfield
centered on the 20~\textmu{}m straylight hole~\textsf{F}. In the blue
we used 256$\times$256 pixels and in the red 128$\times$128 pixels,
the different sizes are necessary because of the difference in image
scale. The top row of Fig.~\ref{fig:restored} shows the central part
of this subfield. We can see directly that the chosen focus positions
were not ideal. PD0 is very similar to WB$-$ and PD$+$ to WB0 in the
blue, so instead of six different diversity channels, we have in
reality only four. The focus is somewhere between WB0 and WB$+$ in the
red; it is probably preferable\footnote{We are not aware of any
  studies about optimal distribution of several focus diversities.} to
have one channel close to the focal plane, to constrain the object
estimation.

Careful pre-processing is easier with point-like targets than with
extended scenes, which facilitates inversions. In particular, it is
possible to fine-tune the intensity bias (dark level) correction, and
then make sure the total energy is the same in all focus positions.
Such refined dark-level correction was performed for the purpose of PD
processing. First, all images in a PD set were normalized with respect
to the average intensity in the inner part of the 1~mm target hole
(\textsf{A}). We assumed this to be the best available measure of the
intensity level, insensitive to the blurring of the edge of the hole.
The WB0 images, which were collected in nominal focus, we then
subtracted a dark level measured as the peak (as defined by a Gaussian
fit) of the histogram of image intensity within the PD processing
subfield. This corrects for some of the straylight from
hole~\textsf{A}. For the data from other focus positions, we
subtracted a dark level corresponding to the difference in median
intensity over the entire frame (defining the dark level). Finally,
the images were normalized to the same mean value, making the total
energy the same in all focus positions.
The dark corrections, $d$, in this refinement step were on the order
$\sim$$10^{-4}$ of the intensity in hole~\textsf{A} in the blue and
almost $\sim$$10^{-3}$ in the red.

In the PD processing the wavefronts were expanded in atmospheric
Karhunen--Lo\`eve (KL) functions, expressed as linear combinations of
Zernike polynomials \citep{roddier90atmospheric}. These KL functions
are ordered as the dominating Zernike polynomial in the notation of
\citet{noll76zernike}, and not by monotonically decreasing atmospheric
variance. Although we were not measuring atmospheric wavefronts, we
chose to use KL functions because bimorph mirrors naturally produce
KL modes.
We assumed that the alignment of the PD channels described in
Sect.~\ref{sec:29maydata} is sufficient for PD processing, and
accordingly did not include tilt coefficients in the fit. We defined
$M$ as the index of the highest-order mode used, i.e., we used KL
modes 4--$M$.

In addition to the KL parametrization of the unknown wavefront, we
also estimated the (Zernike) focus diversities with respect to the WB0
images, starting from the nominal values. Because the WB and PD
cameras sit on a common mount and are therefore moved together, a
single additional focus shift had to be determined when the three
images from the PD camera were added. The diversities estimated with
different $M$ vary by no more than a few tenths of a mm and we used
the values estimated with $M=210$ for both $M$ used here. The nominal
and estimated diversities are shown in Table~\ref{tab:diversity}.

The magnitudes of the estimated diversities are all lower than their
nominal values. This could be caused by a mismatch in the image
formation model. We used the diameter of the ordinary telescope pupil,
re-imaged on the bimorph mirror. In reality, there is diffraction in
the pinhole, making the re-imaged pupil slightly fuzzy. The 7~mm
defocus of the PD camera is particularly underestimated. This may be
because of a mistake in the setup, so this camera was in fact
defocused by less than 7~mm. These estimates confirm that the
diversity setting of the PD camera was not very different from the
step size of the $(-,0,+)$ positions.

\begin{table}[!t]
  \centering
  \caption{Focus diversities.}
  \begin{tabular}{lllllll}
    \hline
    \hline\noalign{\smallskip}
    \multirow{2}{*}{Diversity}&\multicolumn{3}{c}{Blue}&&\multicolumn{2}{c}{Red}\\
    \cline{2-4}\cline{6-7}\noalign{\smallskip}
    & WB$-$ & WB$+$ & PD0 && WB$-$ & WB$+$\\
    \hline\noalign{\smallskip}
    Nominal  (mm)  & $-4.0$  & $+4.0$ & $+7.0$ && $-7.0$  & $+7.0$\\
    Estimated  (mm) & $-3.8$  & $+3.7$ & $+4.8$ && $-6.2$  & $+5.4$\\
    \hline
  \end{tabular}
  \tablefoot{Diversities are expressed in mm shift along the optical
    axis, relative to WB0 in each beam. The PD$\pm$ diversities in the
    blue are simply the sums of the WB$\pm$ and PD0 diversities.}  
  \label{tab:diversity}
\end{table}

\begin{table}[!t]
  \centering
  \caption{RMS and Strehl ratio of the wavefront phase.}
  \begin{tabular}{lrlllllllll}
    \hline
    \hline\noalign{\smallskip}
    \multirow{2}{*}{Beam} 
    & \multirow{2}{*}{$M$}&
    \multicolumn{1}{c}{Measured}
    &&\multicolumn{5}{c}{In focus}
    \\
    \cline{3-3}\cline{5-9}
    \noalign{\smallskip}
    && $\sigma_{\lambda_\text{obs}}$ 
    && $\sigma_{\lambda_\text{obs}}$ 
    & \multirow{2}{*}{$R_{\lambda_\text{obs}}$}
    && $\sigma_{500}$
    & \multirow{2}{*}{$R_{500}$} \\
    &&(waves)
    &&(waves)
    &&&(waves)
    \\
    \hline\noalign{\smallskip}
    Blue & 36  & 0.059  && 0.059  & 0.87 && 0.047 & 0.92 \\
    Blue & 231 & 0.105  && 0.103  & 0.66 && 0.082 & 0.77 \\
    Red  & 36  & 0.100  && 0.052  & 0.90 && 0.065 & 0.85 \\
    Red  & 231 & 0.121  && 0.070  & 0.82 && 0.088 & 0.74 \\
    \hline
  \end{tabular}
  \tablefoot{$\sigma_{\lambda_\text{obs}}$ is the RMS of the PD-estimated
    wavefront in waves at the observations wavelengths, given
    for the entire estimated wavefront phase, as well as after
    subtracting the best-fit Zernike focus component.  
    Strehl ratio calculated as $R_\lambda=\exp\{-(2\pi\sigma_\lambda)^2\}$. 
    Subscript 500 denotes $\lambda=500$~nm.}  
  \label{tab:wfrms}
\end{table}

\begin{figure}[!t]
  \begin{minipage}[c]{\linewidth}
    \centering
    \def\tilewidth{3cm}
    \def\tilewidth{0.24\linewidth}
    \subfloat[Blue 36\label{fig:est-WFblue36}]{\includegraphics[width=\tilewidth]{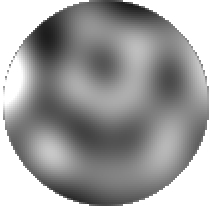}}\hfill
    \subfloat[Blue 231\label{fig:est-WFblue231}]{\includegraphics[width=\tilewidth]{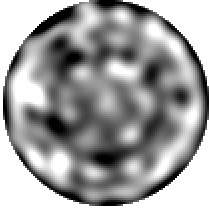}}\hfill
    \subfloat[Red 36\label{fig:est-WFred36}]{\includegraphics[width=\tilewidth]{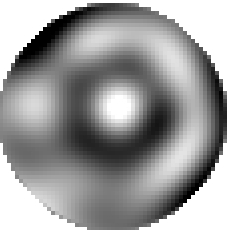}}\hfill
    \subfloat[Red 231\label{fig:est-WFred231}]{\includegraphics[width=\tilewidth]{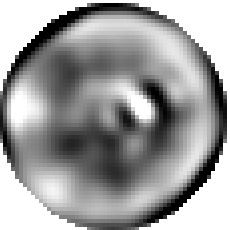}}
    \caption{Estimated wavefronts, $\hat\phi$, in focus. Beam and $M$
      as noted in subcaptions a)--d). The scaling is between the same
      min and max for the two wavefronts from the same beam.}
    \label{fig:est-WF}
 \end{minipage}
\end{figure}

Along with the observed images, $g$, in Fig.~\ref{fig:restored} we
also show the recreated images, $\hat g=\hat f * \hat s_{\phi}$, based
on the PD estimates $\hat f$ and $\hat\phi$, for different $M$. An
inspection of these images clearly shows that the PSFs must be
satisfactorily estimated, particularly in the blue $M=231$ case.

The estimated wavefront phases without focus are shown in
Fig.~\ref{fig:est-WF}. The $M=231$ estimates in the blue and red show
similarities, although the red wavefronts are less well resolved than
the blue wavefronts because of the smaller subfields used and possibly
because of the smaller number of images with different diversities.
They both have two bright rings and a dark gradient at perimeter. Both
rings have brighter bumps at approximately the same positions.
Particularly in the blue wavefront, the pattern resembles that of the
electrodes on the bimorph mirror.
The $M=36$ wavefront estimates are much less resolved. Some
similarities to the 231-estimates can be seen along the outer bright
ring. However, with the smaller $M$, the KL functions apparently
cannot represent the inner ring pattern visible in the $M=231$
wavefront estimates.

The RMS values of the estimated wavefronts are listed in
Table~\ref{tab:wfrms}. The WB camera in the ``0'' position was out of
focus by approximately 0.1~rad in the blue and 0.5~rad in the red. To
make the RMS comparable between beams, we also show the RMS after
subtracting the best-fit focus contribution as well as calculated for
a common wavelength, 500~nm. This makes the 231-estimates in red and
blue agree, $\sigma_{500}=0.085\pm0.003$ waves. The 36 and
231-estimates do not agree in either beam. We are more confident in
the blue 231-results because 1) the results agree between blue and
red, 2) the details in the wavefront are the solution converged to
when $M$ is increased in increments from 36 to 231, and 3) the
solution give estimated PSFs that recreate the observed data well, see
particularly the large-diversity images in Fig.~\ref{fig:restored}.
The Strehl ratio for this solution (see Table~\ref{tab:wfrms}) is 0.66
in the blue. Taking the two 231 solutions in red and blue together we
obtain a Strehl ratio\footnote{The Strehl ratio is the observed peak
  of the PSF divided by the peak of the theoretical PSF for a perfect
  imaging system.} of $75\pm2$\% at 500~nm after removing the focus
error.

Instrumental contributions to the wavefronts can in principle be
estimated and corrected in solar data with the normal MOMFBD image
restoration, except that we usually use 36 modes, far less than 231.
The instrumental high-order modes are not included in the $r_0$
measurements of the wide-field wavefront sensor so they are also not
dealt with in the high-order compensation of
\citet{scharmer10high-order}. They are therefore an independent
contribution to the lowered contrast in our solar images.

\begin{figure}[!t]
  \centering
  \includegraphics[bb=16 63 442 255,width=\linewidth]{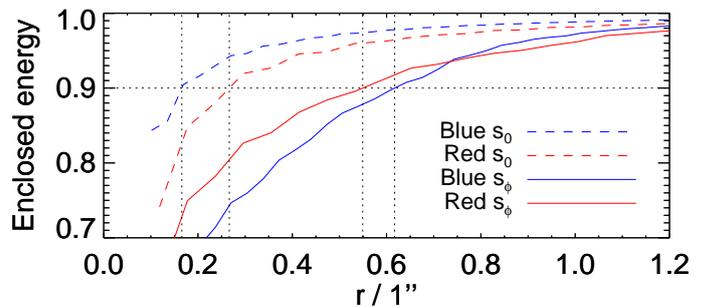}
  \caption{Enclosed energy for the diffraction-limited PSFs, $s_0$,
    and for the estimated wavefront aberration PSFs, $s_\phi$. The
    dotted lines illustrate the radii for 90\% enclosed energy.}
  \label{fig:energy}
\end{figure}
In Fig.~\ref{fig:energy} we show the enclosed PSF energy as a function
of the radial coordinate. The enclosed energies for $s_0$ and $s_\phi$
reach 90\% at $r=0\farcs17$ and $0\farcs62$, respectively, in the blue
and at $0\farcs27$ and $0\farcs55$ in the red.

\subsection{Extended wings}
\label{sec:scattered-light}

\begin{figure}[!t]
  \centering
  \def\tilewidth{0.49\linewidth}
  \reflectbox{\includegraphics[width=\tilewidth]{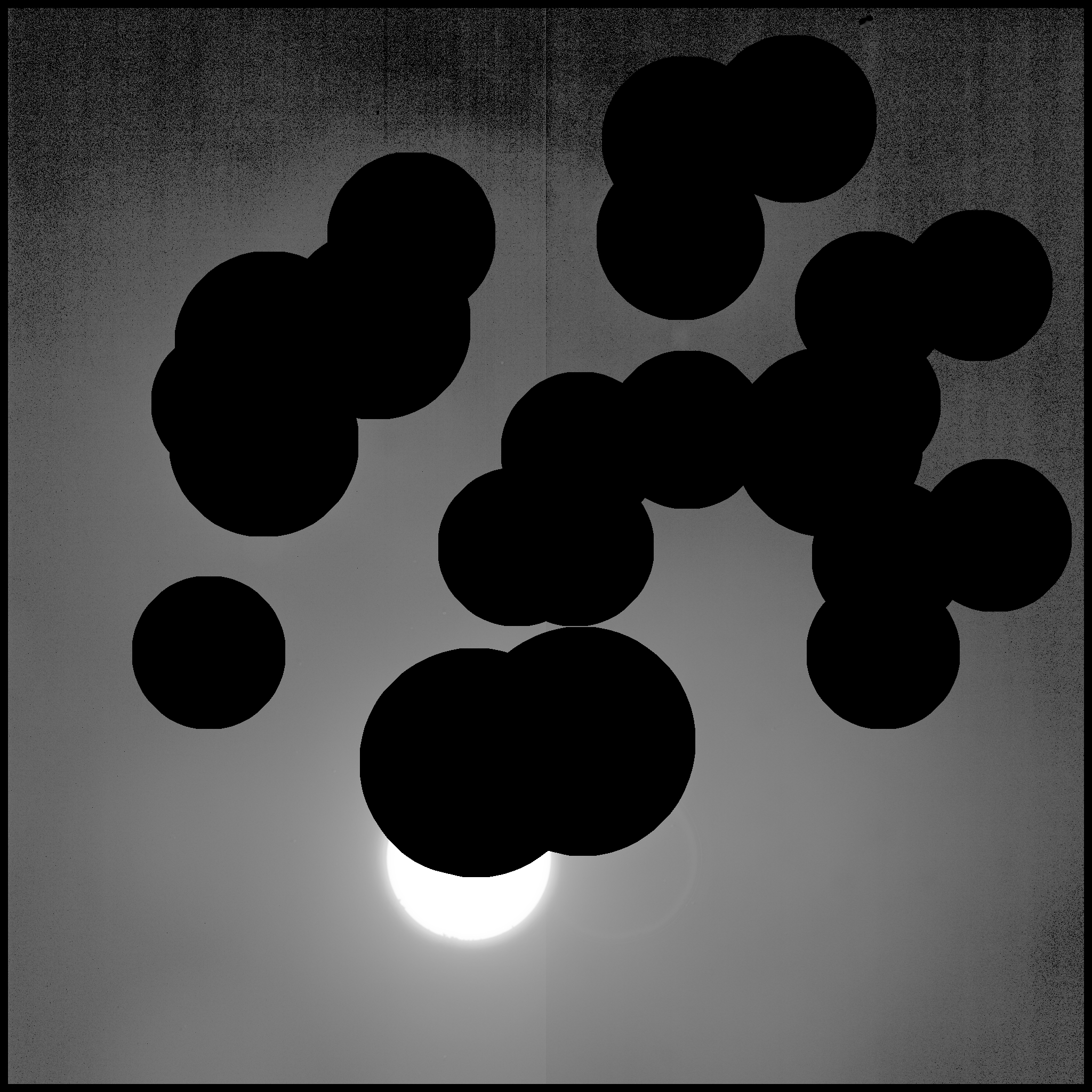}}\hfill
  \reflectbox{\includegraphics[width=\tilewidth]{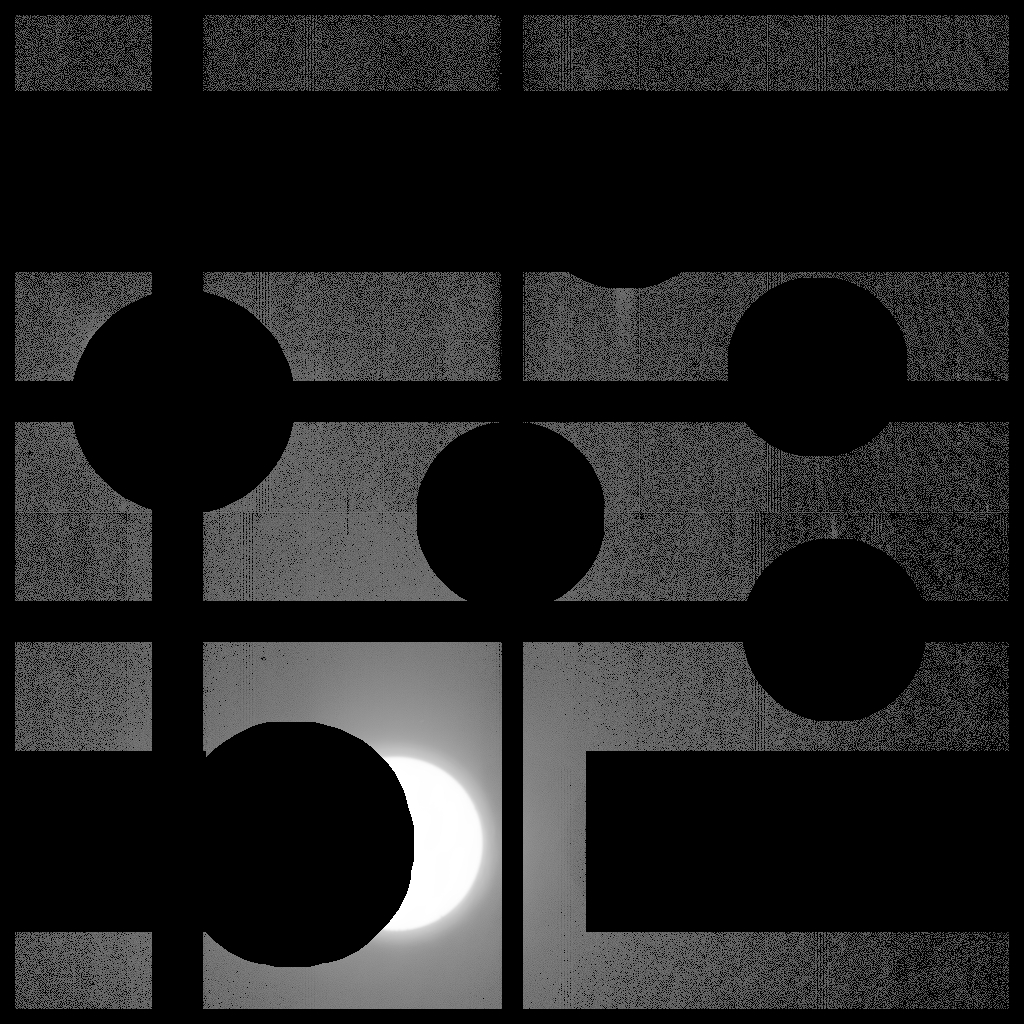}}
  \caption{Masked straylight target images, displayed with the same
    log scale as in Fig.~\ref{fig:images}. \textit{Left}: blue beam;
    \textit{right}: red beam.}
  \label{fig:masked_images}
\end{figure}

We attempted to fit a scattering PSF with extended wings to the WB0
images shown in Fig.~\ref{fig:images}. We preferred to use the 1~mm
hole~\textsf{A}, because it transmits more light than the other holes
and therefore has better SNR in the far wings. To remove the pollution
from the other holes and from the ghost images, we defined binary masks
to be used when fitting, see Fig.~\ref{fig:masked_images}.

For the WB0 images used for fitting the straylight model, the
second-pass sums were corrected for a pattern of stripes running in
the $y$ direction. This correction was made by subtracting from all
rows a smoothed version of the mean of rows 1--10. The average of this
correction was $\sim$$10^{-4}$ in the blue and almost $\sim$$10^{-3}$
in the red, representing a correction of the dark level similar to the
levels corrected in the previous section. (Fig.~\ref{fig:images} shows
the images after this step.)
For this step the images were also normalized with respect to the
average intensity in the inner part of the 1~mm hole~(\textsf{A}).

A binary representation of hole~\textsf{A} was generated by
thresholding the WB0 image at 50\% of the maximum intensity and
removing contributions from the other holes. Artificial images were
then made by convolving this binary image with the appropriate PSFs.

\begin{figure*}[!t]
  \centering \def\tilewidth{0.48\linewidth}
  \includegraphics[bb=29 11 480 337,width=\tilewidth]{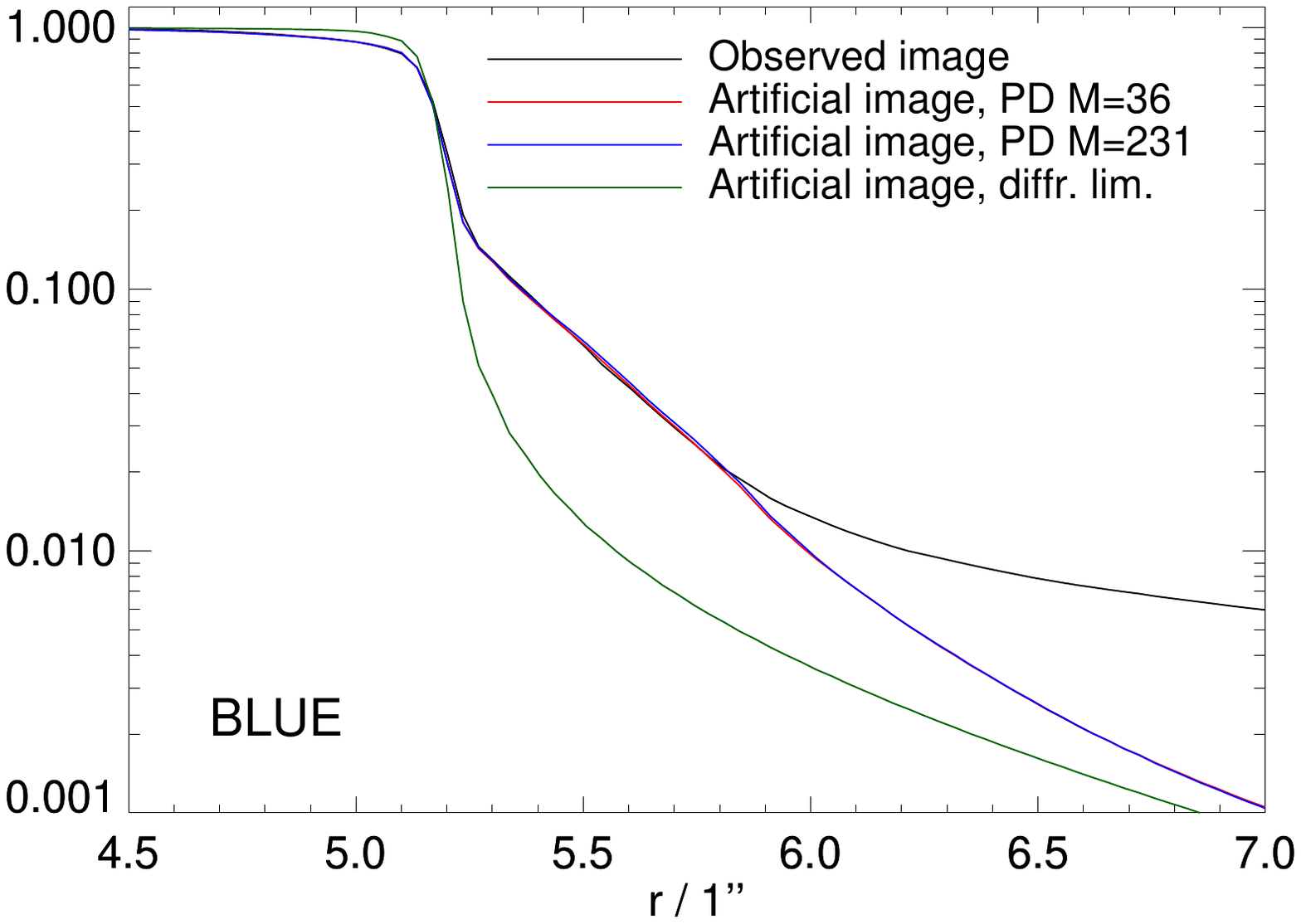}
  \quad
  \includegraphics[bb=29 11 480 337,width=\tilewidth]{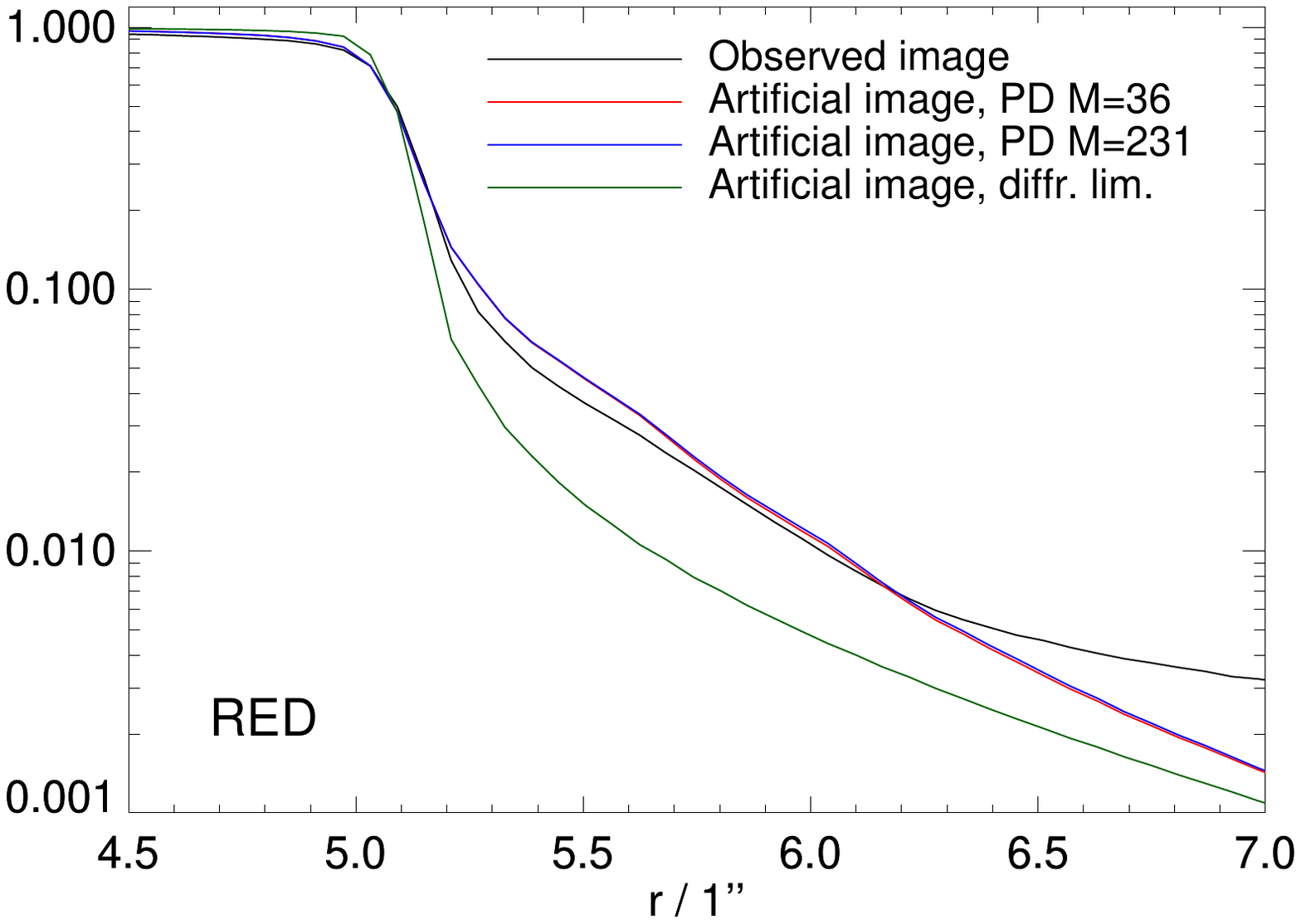}
  \caption{Observed and artificial images, binary 1~mm hole convolved
    with $s_0$ and $s_\phi$ as measured with two different $M$.
    \textit{Left}: blue beam; \textit{right}: red beam.}
  \label{fig:2Dartificial}
\end{figure*}

Figure \ref{fig:2Dartificial} shows an azimuthal average (taking the
mask into account) of hole \textsf{A}, zoomed in on the hole
perimeter. (The zoom makes a discrepancy in image scale of about 1\%
apparent.) We show the artificial hole convolved with three different
PSFs and the observed hole. The artificial image made with PD
estimated $s_\phi$ match the observed data very well in the blue.
Comparing with the diffraction limited image, it is obvious that we
need to model the wavefront PSF to isolate other sources of
straylight. The match is fairly good also in the red but the near
wings are overestimated. The two PD estimated PSFs ($M=36$ and
$M=231$, resp.) give almost indistinguishable results for this
purpose. The straylight we will try to model with the scattering PSF,
$s_K$, is the component that shows up at $r\ga5\farcs8$ in the blue
and $r\ga6\farcs2$ in the red.

We now fitted Eqs.~(\ref{eq:1}) and~(\ref{eq:2}) to the data, using
the $M=231$ estimates of $s_\phi$ as fixed contributions from the
wavefronts. We used the Levenberg--Marquardt algorithm as implemented
in the MPFIT package for IDL
\citep{more78levenberg,2009ASPC..411..251M}. In practice we fitted
with two parameters $c_1$ and $c_2$ substituted for $c$ and $(1-c)$ in
Eq.~(\ref{eq:2}) to allow for normalization differences in real and
artificial data. We then report $c=c_1/(c_1+c_2)$ as the result of the
fit.

\begin{figure*}[!t]
  \centering
  \def\tilewidth{0.48\linewidth}
  \includegraphics[bb=29 11 480 337,width=\tilewidth]{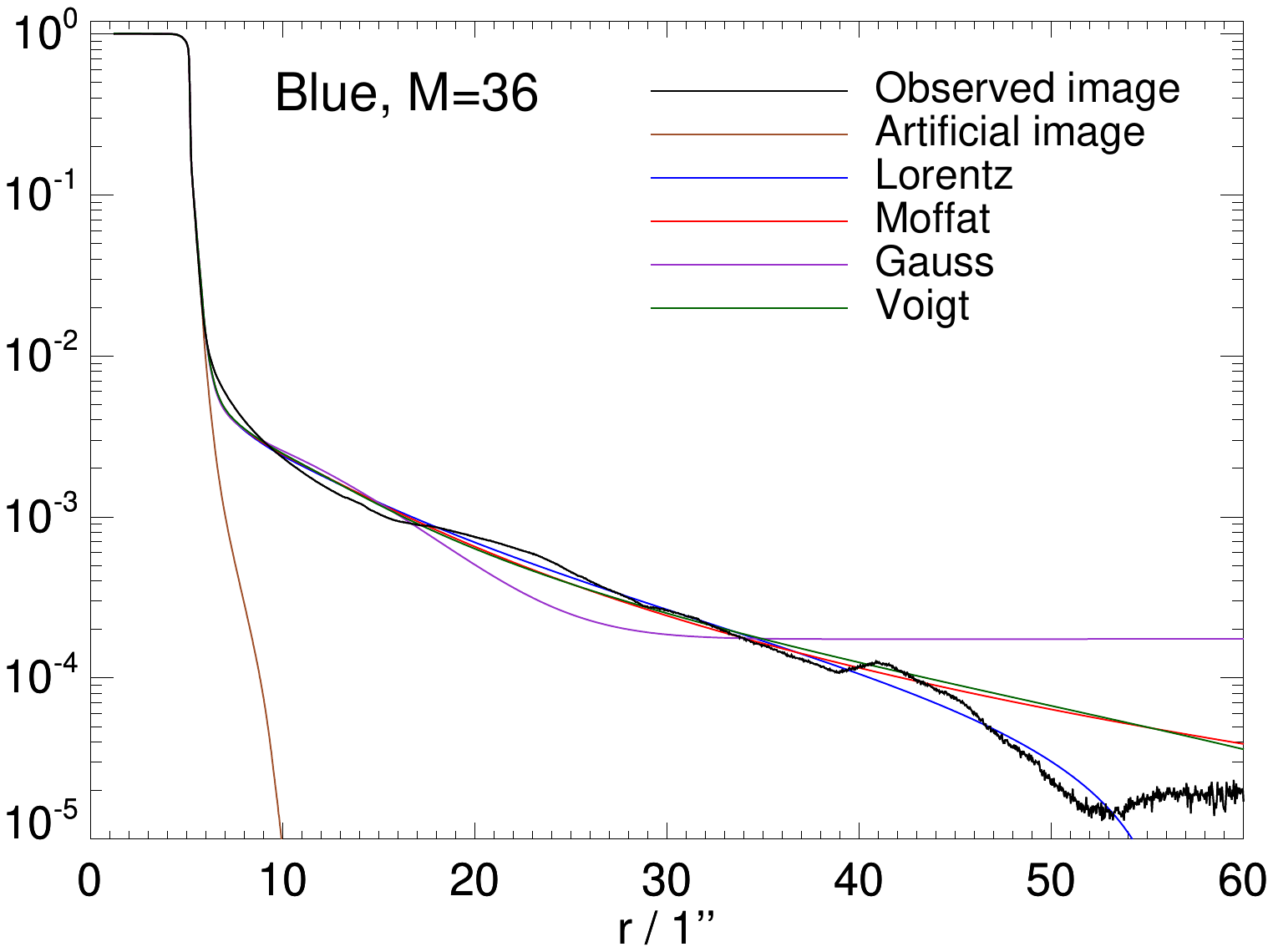}\quad
  \includegraphics[bb=29 11 480 337,width=\tilewidth]{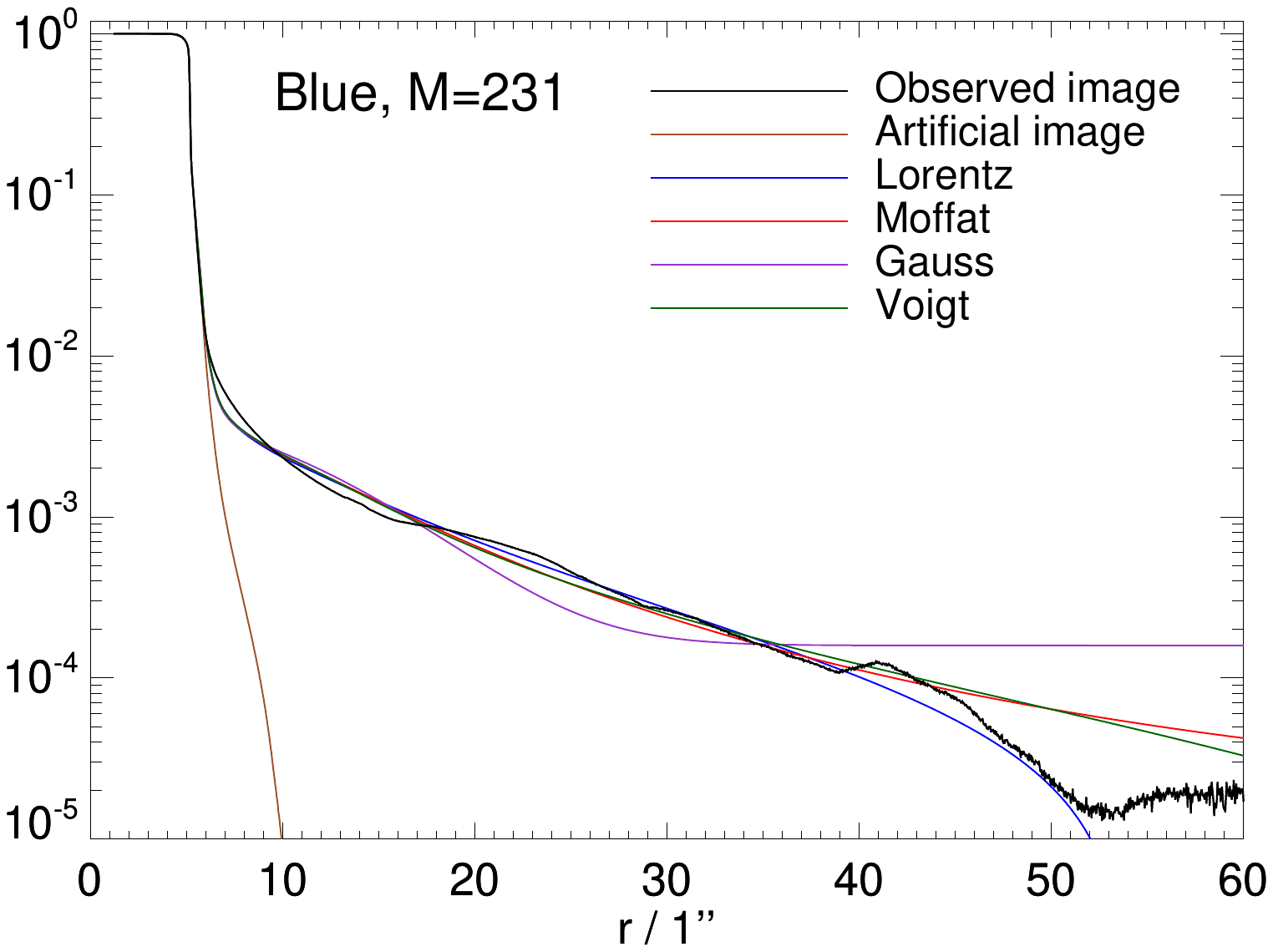}\\[3mm]
  \includegraphics[bb=29 11 480 337,width=\tilewidth]{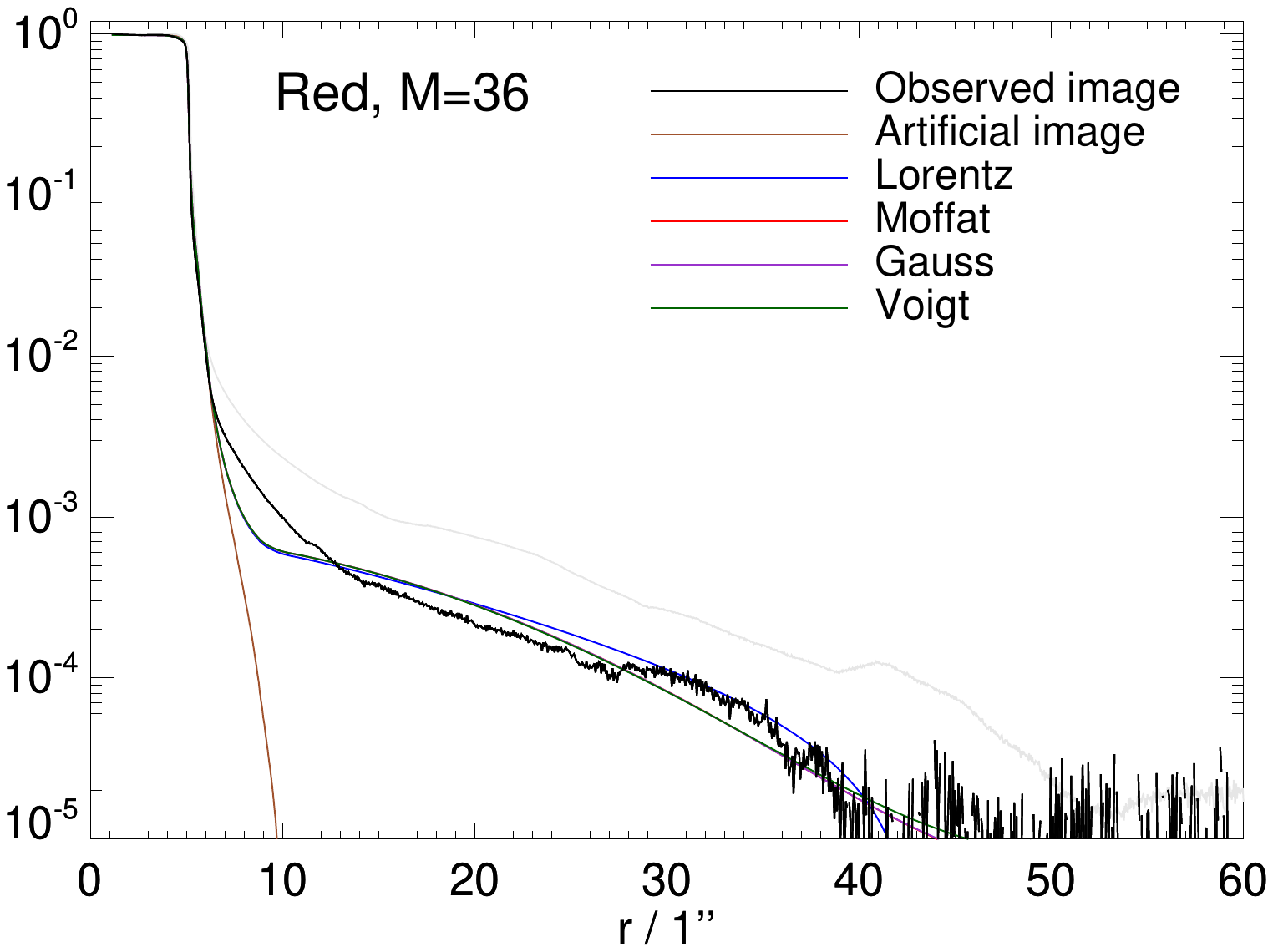}\quad
  \includegraphics[bb=29 11 480 337,width=\tilewidth]{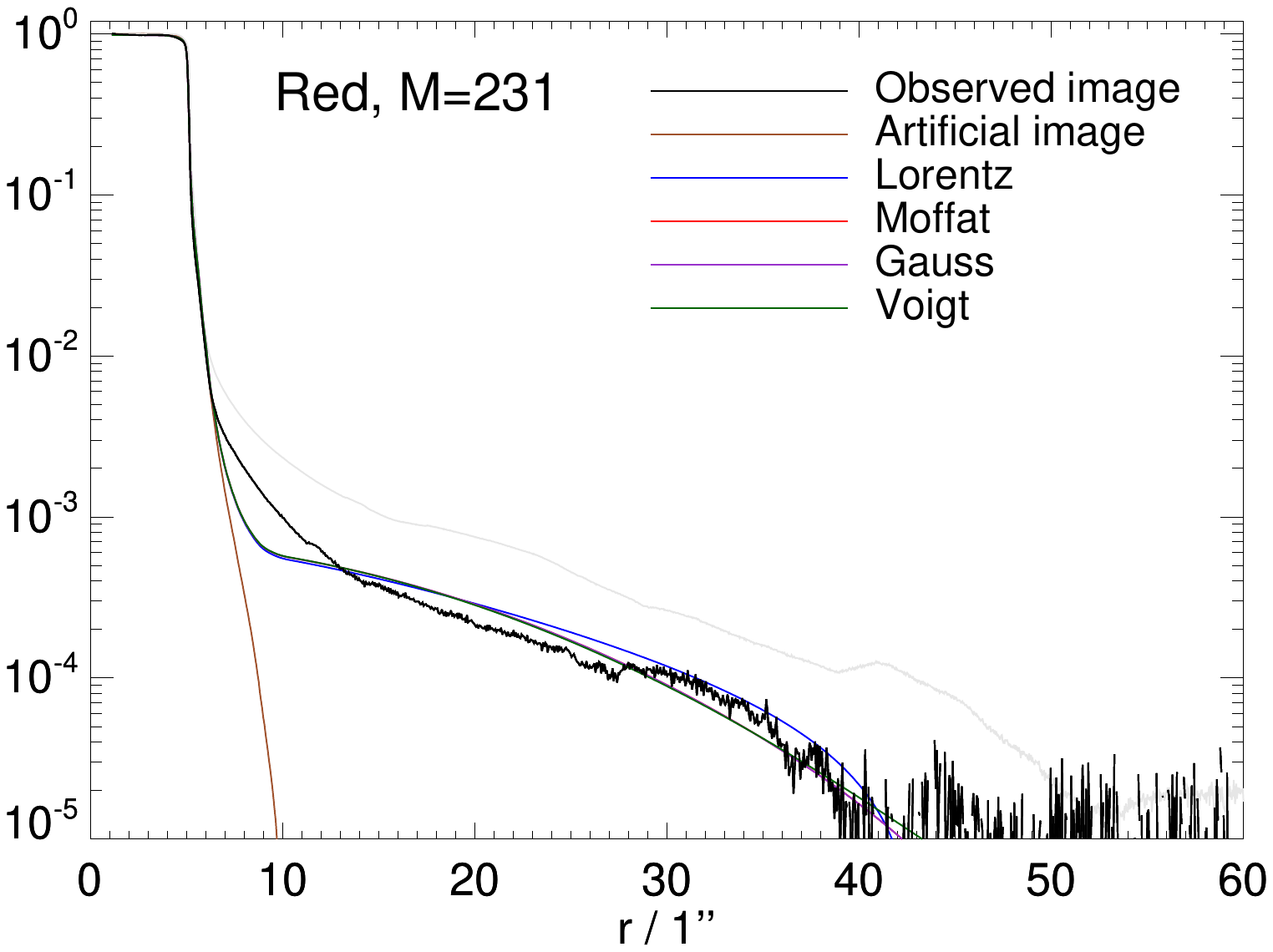}
  \caption{Fit results, angular averages of the masked images,
    centered on the 1 mm hole \textsf{A}. Beam and $M$ as noted in the
    plot legends. Black: observed image; gray (in the red plots): blue
    observed image; brown: binary image convolved with $s_\phi$
    estimated with $M=231$; Other colors: binary image convolved with
    $s_\phi$ as well as $s_K$ fitted with kernels as listed.}
  \label{fig:1DFits}
\end{figure*}

\begin{table*}[!t]
  \def\0{\phantom0}
  \centering
  \caption{Scattering kernel fit results}
  \label{tab:Scatteringfits}
  \begin{tabular}{rllrrllllrrll}
    \hline
    \hline\noalign{\smallskip}
    \multirow{2}{*}{$M$} & \multirow{2}{*}{Kernel} & \multicolumn{5}{c}{Blue} && \multicolumn{5}{c}{Red}\\
    \cline{3-7}\cline{9-13}\noalign{\smallskip}
    && $c$  & \multicolumn{2}{l}{Kernel parameters} & FWHM & $\chi^2$ &&  $c$  & \multicolumn{2}{l}{Kernel parameters} & FWHM & $\chi^2$\\
    \hline\noalign{\smallskip}
     36 & Gauss   & 99.8 & $\sigma=\08.31$ &                & 19\farcs6 & 0.0143 && 99.9 & $\sigma=\013.74$ &                  & 32\farcs4 & 0.0595 \\
     36 & Lorentz & 99.6 & $\gamma=\06.61$ &                & 13\farcs2 & 0.0142 && 99.7 & $\gamma=\017.78$ &                  & 35\farcs6 & 0.0595 \\
     36 & Moffat  & 99.7 & $\alpha=\09.03$ & $\beta=  1.36$ & 14\farcs7 & 0.0142 && 99.9 & $\alpha= 416.03$ & $\beta= 458.67$  & 32\farcs4 & 0.0595 \\
     36 & Voigt   & 99.7 & $\sigma=\03.98$ & $\gamma= 4.23$ & 14\farcs7 & 0.0141 && 99.9 & $\sigma=\013.64$ & $\gamma= 0.0008$ & 32\farcs1 & 0.0595 \\
    231 & Gauss   & 99.8 & $\sigma=\08.77$ &                & 20\farcs6 & 0.0148 && 99.9 & $\sigma=\014.61$ &                  & 34\farcs4 & 0.0593 \\
    231 & Lorentz & 99.6 & $\gamma=\07.38$ &                & 14\farcs8 & 0.0147 && 99.6 & $\gamma=\019.83$ &                  & 39\farcs7 & 0.0593 \\
    231 & Moffat  & 99.7 & $\alpha= 11.04$ & $\beta=  1.56$ & 16\farcs5 & 0.0147 && 99.9 & $\alpha= 524.28$ & $\beta= 644.53$  & 34\farcs4 & 0.0593 \\
    231 & Voigt   & 99.7 & $\sigma=\04.59$ & $\gamma= 4.26$ & 16\farcs1 & 0.0147 && 99.9 & $\sigma=\014.39$ & $\gamma= 0.0008$ & 33\farcs9 & 0.0593 \\
    \hline
  \end{tabular}
  \tablefoot{$M$ is the max KL index used in PD. The parameters
    $\alpha$, $\gamma$, and $\sigma$ are in units of arcsec, while
    $\beta$ is dimensionless. The $c$ parameter is the
    percentage of the intensity that is not contained within the far wings.}
\end{table*}

The results of the fits are shown in Fig.~\ref{fig:1DFits} and
in Table~\ref{tab:Scatteringfits}. With any of the kernels we get
$c>0.99$ in both beams, which appears to be a very robust result by
the agreement between the different fits. In the blue we get
0.2--0.4\% scattered light and in the red 0.1--0.4\%.

The $d$ parameter is returned with low values, on the order
of~$10^{-6}$. This indicates that the subtraction of the average top
rows worked well as a dark correction refinement.
Note that in the red, outside the signal-dominated radii, there is
just noise, seemingly around zero. In the blue there seems to be a
remaining uncorrected dark level that is not modeled by the $d$ fit
parameter. This could be caused by a geometrical asymmetry, e.g., by
insufficient masking of the other holes or a dark-level gradient. It
does not seem to ruin the fits.

We have less confidence in the details of the red fits than in the
blue fits because the artificial image based on $\hat s_{\phi}$
overestimates the core of the PSF, see the right panel of
Fig.~\ref{fig:2Dartificial}. However, it is apparent from
Fig.~\ref{fig:1DFits} that there is less straylight in the red than in
the blue. Comparison of the observed red data (black in the red plots)
with the observed blue data (gray) reveals a factor of $\sim$3
difference. This would favor the lower estimate of 0.1\% in the red
and make the Lorentz kernel fit less believable.

In the blue the fit with a Gaussian kernel seems to fail at large
radii. It also has the largest FWHM, $\sim$$20\arcsec$. The Lorentzian
kernel follows the signal dominated curve better than the other
kernels but the Voigt and Moffat kernels give similar results. These
all have FWHM 13\arcsec--16\arcsec.
In the red the Lorentzian kernel differs from the other kernels by
giving a larger FWHM. While the Moffat kernel $\beta$ is estimated
with a very high value, which is compensated for by an $\alpha$ that
is also very large, and the Voigt kernel is almost degenerated to a
Gaussian, the latter three kernels agree on a FWHM of 34\arcsec.
We conclude that the Voigt and Moffat kernels, by virtue of their two
fit parameters, are better able to represent the true shape of the
extended wings. Using their results, we estimate the FWHM to
16\arcsec{} in the blue and 34\arcsec{} in the red.

There are noticeable residuals at $10\arcsec\pm\text{ a few arcsec}$
in the fits shown in Fig.~\ref{fig:1DFits}. Particularly in the red,
where this feature corresponds to 36\% of the total intensity! Because
of the log scale, this residual looks much lower in the blue but the
corresponding error in the blue is in fact 15\%. This may sound
alarming but it appears this is needed to cancel fitting errors within
the hole and at the hole perimeter.
There are several possible reasons for the bad fits at small radii.
There are small imperfections in the shape of the hole that could
cause diffraction. The 50\% threshold does not necessarily produce a
binary representation of the hole with a correct size. The thickness
of the metal foil may cause reflections in the interior walls of the
hole. There are remaining variations in intensity of unknown origin
within the hole. Paricularly in the red, the PD estimated $s_\phi$
does not fit the core of the PSF. Owing to the truncation of the
wavefront expansion, the estimated $s_\phi$ may also under-represent
the near wings of the PSFs. Thermal relaxation may cause variations in
the $s_\phi$ errors around the hole perimeter during the data
collection. Despite the limited accuracy at small radii, the residuals
for the extended wings at larger radii are low and we believe the
kernel fits can be trusted because the 2D fitting is dominated by the
larger areas where the radii are large.

\section{Discussion}
\label{sec:discussion}

The wavefront aberrations are quite significant, corresponding to a
Strehl ratio of $\sim$75\%. The structure of the estimated wavefronts
($M=231$) with their 6- and 12-fold symmetries suggests that the
bimorph mirror of the SST AO is responsible. Wavefront aberrations
originating in the instrumentation will be at least partly corrected
for by our standard MOMFBD image restoration. Part of the wavefront
aberrations came from modes of higher order than we normally include
in this processing and it is unclear how much of this would be
corrected for. With the point-like object used here, the magnitude of
the estimated wavefront strongly depends on the number of included
modes, $M$. However, as shown by \citet{scharmer10high-order}, when
the object is solar granulation, the MFBD/PD-type problem is less
constrained and an estimated low-order wavefront tends to also
represent the blurring caused by the higher-order modes that are not
included.

The aberrations give straylight that is mostly contained in the first
diffraction rings of $s_\phi$, 90\% of the energy is within a radius
of 0\farcs6 (see Fig.~\ref{fig:energy}). The origin of this
surprisingly high level of wavefront errors is unknown. One possible
source is the mounting of the 1~mm thick deformable mirror, which is
clamped between two O-rings with the tension adjusted manually by 12
screws. Another possibility is high-order aberrations induced by a
large focus error imposed on the mirror.

\citet[see the supporting online material]{scharmer11detection}
compared umbra intensity and granulation contrast in 630~nm SST data
to data from SOT/Hinode and inferred a 50\% straylight level with a
small FWHM of less than 2\arcsec{}, consistent with this straylight
originating from aberrations. However, the aberrations measured here
can only account for about 1/3 of the 50\% straylight and would have
to be stronger by a factor 2 than those measured here  to
fully explain the contrast in the observed granulation data.

Our best estimate of the scattering PSF, $s_K$, is that it generates
only 0.3\% straylight in the blue ($\sim$397~nm) and 0.1\% in the red
($\sim$630~nm). In our best fits, the FWHM of the scattering kernel
$K$ is $\sim$16\arcsec{} in the blue and $\sim$34\arcsec{} in the red,
using the Voigt and Moffat kernels. The Gauss and Lorentz kernel fits
give similar results.

It is common to truncate the wings of the scattering kernels at some
radius but our fits work well without truncation, possibly because we
are including the dark level $d$ in the fit. 
%
We cannot exclude that some pollution from the holes
\textsf{B}--\textsf{F} influenced the $s_K$ fits. A better straylight
target should have only the 1~mm hole and the smallest hole.

After the ordinary dark-current calibrations, which consist of a
subtraction of an average dark frame, the dark level was manipulated
in various ways. For PD processing, we estimated and subtracted
$\sim$$10^{-4}$ of the intensity in hole \textsf{A} from the subfield
containing hole \textsf{F}, measured basically as the modal value
within the PD processing subfield. Before the straylight kernel
fitting, we also subtracted on the order of $\sim$$10^{-4}$ based on
the top 10 rows of pixels. Because the kernel fitting involved a dark
level, $d$, which was estimated to $\sim$$10^{-6}$, we can be
reasonably sure that the standard dark-correction is correct to the
$10^{-4}$ level. However, this is insignificant in comparison with the
ghost images, the most significant of which could be measured to
$\sim$1\% of the hole \textsf{A} intensity. Consequently, we have to
allow for an additive component at the $\sim$$10^{-2}$ level, which is
not modeled by the scattering kernels.

\section{Conclusions} 
\label{sec:conclusions}

We have proposed a procedure for measuring the amount of straylight in
the SST post-focus instrumentation and applied it to one wavelength in
the blue beam and one wavelength in the red beam. The strength of the
method is that we simultaneously measured wavefront aberrations (with
phase diversity methods) and ``conventional'' scattered light from the
same data. Thus, we were able to separate these two sources of straylight.

The dominant contributions to straylight are high-order aberrations,
causing a reduction of the Strehl ratio. The estimated Strehl ratio at
390, 500 and 630~nm is 66\%, 75\%, and 82\%, resp., corresponding to
integrated straylight within a radius of $\sim$0\farcs6 on the order
of 34\%, 25\%, and 18\%, resp. The second-most important source of
straylight is multiple weak, out-of-focus ghost images. The combined
effect of these is difficult to estimate but most likely contributes
less than a few percent at all wavelengths. The smallest contribution
measured is from ``conventional'' straylight in the form of a PSF with
very wide wings (16\arcsec{} in the blue, 34\arcsec{} in the red),
this is estimated to contribute 0.3\% in the blue beam and 0.1\% in
the red.

The wavefront contribution could potentially be higher when observing
the Sun because of the heat that affects the mirror. Considering the
small number of optical surfaces and the high optical quality of the
SST telescope optics, other main contributions to the lowered contrast
most likely are of atmospheric origin. This could come partly from
uncorrected high-order wavefront aberrations, as discussed by
\citet{scharmer10high-order}, and partly from scattering by dust
particles in the Earth's atmosphere.

\balance
\begin{acknowledgements}       

  We thank Vasco Henriques and Pit S\"utterlin for help with the data
  collection.
  The Swedish 1-m Solar Telescope is operated on the island of La
  Palma by the Institute for Solar Physics of the Royal Swedish
  Academy of Sciences in the Spanish Observatorio del Roque de los
  Muchachos of the Instituto de Astrof\'isica de Canarias.

\end{acknowledgements}

\noindent\emph{Note added in proof}
In October 2011 we were upgrading the wavefront sensor of the SST AO
system. We found that switching off the mirror voltages results in a
change in the science focus by approximately 9~cm along the optical
axis. Due to limited electrode resolution, deformable mirrors cannot
produce even low-order modes perfectly. When such a large focus has to
be compensated for, there are therefore unavoidable high-order
wavefront errors. We have simulated our setup and calculated the
wavefront phase corresponding to the difference between a 9~cm defocus
and the approximate focus produced by the mirror in order to
compensate for it. The resulting RMS wavefront is 0.14~waves at 500~nm
and the corresponding Strehl ratio is 0.48. This is even worse than
the effects reported in Table~\ref{tab:wfrms} by a factor 1.6 in RMS
wavefront.

We note again that these measurements (and simulations) are not
directly comparable to solar observations. Nevertheless, this effect
may well contribute by a significant amount to the loss of contrast in
SST data. The SST optical setup will be modified to take this into
account when we install our new 85-electrode deformable mirror during
the summer or fall of 2012. Before that, we will also try to
compensate for the effect by introducing a corresponding focus change
in the Schupmann system.


\end{document}